\documentclass[prb,a4paper,showpacs,showkeys,twocolumn,amsmath,amssymb,citeautoscript,floatfix]{revtex4}

\preprint{$ $Id: revtexpaper.tex,v 1.31 2005/04/07 14:19:24 ben Exp $ $}

\usepackage{graphicx}
\usepackage{bm,amsmath}
\DeclareGraphicsExtensions{.eps}
\usepackage{hyperref}

\newcommand{\Equation}[1]{Eq.~\ref{#1}}
\newcommand{\Figure}[1]{Fig.~\ref{#1}}
\newcommand{\Section}[1]{\S\ref{#1}}

\newcommand{\me}{\mathrm{e}} 
\newcommand{\mi}{\mathrm{i}} 
\newcommand{\dif}{\mathrm{d}} 
\renewcommand{\vec}[1]{\mathbf{#1}} 
\newcommand{\unitvec}[1]{\hat{\vec{#1}}} 

\newcommand{\abs}[1]{\left\lvert#1\right\rvert} 
\newcommand{\defined}{\stackrel{\text{def}}{=}} 

\DeclareMathOperator{\sgn}{sgn} 
\DeclareMathOperator{\curl}{curl} 
\newcommand{\scp}{\bm{\cdot}} 

\newcommand{\bragg}{\theta_{\text{B}}} 

\newcommand{\elsus}{\ensuremath{\chi}} 
\newcommand{\strain}{\ensuremath{\epsilon}} 

\newcommand{\cm}{\,\text{cm}}
\newcommand{\um}{\,\mu\text{m}}
\newcommand{\nm}{\,\text{nm}}

\newcommand{\us}{\,\mu\text{s}}
\newcommand{\ns}{\,\text{ns}}
\newcommand{\ps}{\,\text{ps}}
\newcommand{\fs}{\,\text{fs}}

\newcommand{\mps}{\,\text{ms}^{-1}} 

\newcommand{\keV}{\,\text{keV}}
\newcommand{\eV}{\,\text{eV}}

\newcommand{\fluence}{\,\text{mJ}/\text{cm}^{2}}

\begin{document}

\title{Simulations of Time-Resolved X-Ray Diffraction in Laue Geometry}

\date{\today}

\author{B.~Lings}
\affiliation{Department of Physics, Clarendon Laboratory, University of Oxford, Oxford, OX1 3PU, United Kingdom}
\author{M.~F.~DeCamp}
\affiliation{Department of Chemistry, Massachusetts Institute of Technology, Cambridge, Massachusetts 02139, USA}
\author{D.~A.~Reis}
\affiliation{FOCUS Center and Department of Physics, University of Michigan, Ann Arbor, Michigan 48109, USA}
\author{S.~Fahy}
\affiliation{Physics Department and NMRC, University College, Cork, Ireland}
\author{J.~S.~Wark}
\homepage{http://laserplasma.physics.ox.ac.uk/}
\affiliation{Department of Physics, Clarendon Laboratory, University of Oxford, Oxford, OX1 3PU, United Kingdom}

\begin{abstract}
A method of computer simulation of Time-Resolved X-ray Diffraction (TRXD) in asymmetric Laue (transmission) geometry with an arbitrary propagating strain perpendicular to the crystal surface is presented. We present two case studies for possible strain generation by short-pulse laser irradiation: (i) a thermoelastic-like analytic model \cite{thomsen86}; (ii) a numerical model including effects of electron-hole diffusion, Auger recombination, deformation potential and thermal diffusion.  A comparison with recent experimental results~\cite{decamp03} is also presented.
\end{abstract}

\pacs{07.05.Tp, 61.10.-i, 78.47.+p}

\keywords{dynamical diffraction; time-resolved x-ray diffraction}

\maketitle

\section{Introduction}
When a semiconductor crystal is irradiated by photons of above band-gap energy, electrons are excited from the valence band to the conduction band.  If the light source is a subpicosecond laser, the electron-hole plasma density can easily reach $10^{20}\cm^{-3}$---enough for the lattice spacing changes due to the deformation potential to be significant and for the fast diffusion of the electron-hole plasma into the crystal bulk to be appreciable.  The time-scale of the fastest energy transfer processes from the electrons to the lattice is of the order of picoseconds or less, which is significantly shorter than the hydrodynamic response time of the crystal.  The resultant stress is therefore relieved by surface expansion and, by Newton's third law, a bipolar compression wave propagating into the crystal. These effects have been studied by optical methods~\cite{auston74,auston75,chigarev00,tanaka95} and, more recently, as test-cases for the field of Time-Resolved X-ray Diffraction (TRXD) in Bragg (reflection) geometry~\cite{rose-petruck99,lindenberg00,reis01,siders99,cavalleri00,sokolowski-tinten01,cavalleri01}.  However, although x-ray diffraction in Bragg geometry allows the direct study of the structural changes to the lattice---as opposed to the electronic effects in optical studies---the very nature of Bragg geometry x-ray diffraction with perfect crystals only allows the study of the surface region of the crystal.  By utilising the anomalous absorption effect in Laue (transmission) geometry, recent experiments\cite{decamp01,decamp03} have enabled the probing of structural changes throughout the bulk of the crystal on a sub-nanosecond time-scale, giving a much clearer picture of the phenomena under consideration.

In Bragg geometry it is only possible to study the structural changes of crystalline matter in a layer with thickness comparable to the shorter of the x-ray extinction depth ($\sim\um$) for strong diffraction or the absorption depth for weak diffraction~\cite{reis01}.  In Laue geometry it is possible to study the whole bulk of the crystal due to anomalous absorption, known as the Borrmann effect.  In dynamical diffraction theory, the x-ray electromagnetic field inside the crystal is resolved into two independent eigensolutions---one with nodes on the lattice planes, and the other with anti-nodes.  This creates one solution with reduced absorption, and another with enhanced absorption.  These are known respectively as the $\alpha$ and $\beta$ branches. It is the $\alpha$ branch solution that can propagate for many extinction depths without significant loss---probing the entire depth of the crystal~\cite{batterman64}.  A disturbance in the crystal lattice causes energy to be transferred between the two branches, an effect known as inter-branch scattering~\cite{authier01}.  When the x-ray beam exits the crystal, it is again expressed as the free-space solutions: the $0$ (forward-diffracted) and $h$ (deflected-diffracted) beams, which are linear combinations of the $\alpha$ and $\beta$ branches.  If the ratio and/or relative phases of the branches has changed, the energy partitioning between the beams will be affected, often to a significant degree~\cite{decamp01}.

Using this method it is possible to study the mechanism of energy transfer when a single crystal target is irradiated by a femtosecond laser pulse.  We will show that it is possible to distinguish between a model which assumes an instantaneous energy transfer into the lattice, and one which models thermal diffusion, electron-hole plasma diffusion, the effect of the deformation potential, and energy transfer to the lattice via Auger recombination.

\section{Theory}

The fundamental assumption in this method of diffraction simulation is that the strain field in the crystal is one-dimensional, parallel to the surface normal, and can be approximated by a constant strain in many lamin\ae, each one parallel to the surface of the crystal.  The degree to which this represents the experimental situation is discussed in \Section{sec:strain-models}. The electric displacement field amplitude is then calculated as it passes through these lamin\ae, taking into account the effect of the strain in each.

\subsection{Dynamical Diffraction Theory}
\label{sec:dynamical-theory}

\begin{figure}
\begin{center}
\includegraphics[width=\columnwidth]{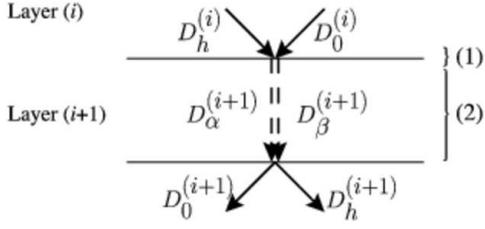}
\caption{Layer approximation. (1) At the boundaries between layers, the $\alpha$ and $\beta$ branches are calculated (\Equation{eq:branches}). (2) The beam is then propagated through the layer (\Equation{eq:propagation}).}
\label{fig:layer-approximation}
\end{center}
\end{figure}

By solving the wave equation
\begin{equation}
\curl\curl\big[(1-\elsus)\vec{D}(\vec{r})\big]=4\pi^2 k^2 \vec{D}(\vec{r})
\end{equation}
in a medium with a periodic susceptibility $$\chi(\vec{r})=\sum_{h}\chi_{h}\me^{-2\pi\mi\vec{h}\scp\vec{r}},$$ using a displacement field $$\vec{D(\vec{r})}=\sum_{h}\vec{D}_{h}\me^{-2\pi\mi\vec{k}_{h}\scp\vec{r}},$$ where $\vec{k}_{h}=\vec{k}_{0}+\vec{h}$ is the wave-vector of the $h$-th component inside the crystal and both are summed over fourier components in the reciprocal lattice vectors $\vec{h}$, we obtain the fundamental equations of dynamical theory~\cite{zachariasen45}
\begin{equation}
\label{eq:fundamental}
\sum_{h'}\chi_{h-h'}\left[(\vec{k}_h\scp\vec{D}_{h'})\vec{k}_h-k_h^2\vec{D}_{h'}\right]
=(k^{2}-k_{h}^{2})\vec{D}_{h}.
\end{equation}
These can be further simplified by the assumption that there are only two strong x-ray wave-fields (incident and diffracted: $\vec{D}(\vec{r})=\vec{D}_{0}\me^{-2\pi\mi\vec{k}_{0}\scp\vec{r}}+\vec{D}_{h}\me^{-2\pi\mi\vec{k}_{h}\scp\vec{r}}$
  ), leading to the two-beam dispersion relation
\begin{subequations}
\label{eq:fundamental-2-beam}
\begin{equation}
\elsus_{\bar{h}}\left[(\vec{k}_0\scp\vec{D}_{h})\vec{k}_0-k_0^2\vec{D}_h\right]
=\left[k^2-k_0^2(1-\elsus_0)\right]\vec{D}_0,
\end{equation}\begin{equation}
\elsus_{h}\left[(\vec{k}_h\scp\vec{D}_{0})\vec{k}_h-k_h^2\vec{D}_0\right]
=\left[k^2-k_h^2(1-\elsus_0)\right]\vec{D}_h.
\end{equation}
\end{subequations}
These vector equations relate the physically realisable values of $\vec{D}_{0,h}$ and can be solved for the incident and diffracted beam amplitude as a function of depth into a crystal layer giving equations of the form (for $\sigma$-polarization)
\begin{equation}
\label{eq:propagation}
\begin{pmatrix}
D_0 \\ D_h
\end{pmatrix}
=
\begin{pmatrix}
\me^{-2\pi\mi\vec{k}_\alpha\scp\vec{r}} &
 \me^{-2\pi\mi\vec{k}_\beta\scp\vec{r}} \\
\xi_\alpha \me^{-2\pi\mi\vec{k}_\alpha\scp\vec{r}} & 
\xi_\beta \me^{-2\pi\mi\vec{k}_\beta\scp\vec{r}}
\end{pmatrix}
\begin{pmatrix}
D_\alpha \\ D_\beta
\end{pmatrix},
\end{equation}
where
\begin{equation}
\vec{k}_\alpha\defined \frac{k\delta_0'\unitvec{n}}{\gamma_0},\quad
\vec{k}_\beta\defined \frac{k\delta_0''\unitvec{n}}{\gamma_0},\quad
\end{equation}
are the wave-vector shifts of the $\alpha$ and $\beta$ branches (normal to the crystal surface---see \Figure{fig:layer-approximation}), $D_{\alpha}$ and $D_{\beta}$ are the amplitudes of the $\alpha$ and $\beta$ branches,
\begin{align}
\left.\begin{array}{l}
\delta_0' \\
\delta_0''
\end{array}
\right\}
&=\tfrac{1}{2}\left[\elsus_0 - z\pm\sqrt{q+z^2}\right],
\\
\left.\begin{array}{l}
\xi_\alpha \\
\xi_\beta
\end{array}
\right\}
&=\frac{-z \pm \sqrt{q+z^2}}{\elsus_{\bar{h}}},
\end{align}
\begin{equation}
z\defined\frac{1-b}{2}\elsus_0 - b \Delta\theta \sin2\bragg,\label{eq:alpha-definition} \quad
q\defined b\elsus_h\elsus_{\bar{h}},
\end{equation}
\begin{equation}
b\approx\frac{\gamma_0}{\gamma_h}, \quad
\Delta\theta = \theta-\bragg,
\end{equation}
where $\gamma_0,\gamma_h$ are the direction cosines of the incoming and outgoing beams respectively, in relation to the surface normal of the crystal $\unitvec{n}$; $\xi_{\alpha,\beta}$ are the amplitude ratios $D_h/D_0$; $\theta$ is the actual glancing angle of incidence relative to the diffracting planes; and $\bragg$ is the Bragg glancing angle\cite{zachariasen45}.  One can also define a dimensionless deviation parameter $\eta$, defined as
\begin{equation}
\label{eq:eta-definition}
\eta=\frac{\Delta\theta - \Delta\theta_{os}}{\delta_{os}}
\end{equation}
where $\Delta\theta_{os}$ is the refraction shift of the diffraction peak and $2\delta_{os}$ is the Darwin width~\cite{authier01}.  This can be shown to be equal to $-\frac{z}{\sqrt{q}}$ in terms of the quantities shown above.  The deviation parameter can also be expressed as a function of wavelength separation from the Bragg wavelegth $\Delta\lambda = \lambda - \lambda_{\text{B}}$, using the relation $\Delta\lambda=(\lambda_{\text{B}}\cot\bragg)\Delta\theta$ derived from Bragg's law.  This is the form that will be used for analysing the experimental data.  However, it is easier to visualise the effects of strain (see the following section) when it is considered as a function of angle.

\Equation{eq:propagation} can be solved at the boundary between two layers ($\unitvec{n}\scp\vec{r}=0$) to find $D_{\alpha}$ and $D_{\beta}$ as functions of $D_{0}$ and $D_{h}$ in the previous layer:
\begin{equation}
\label{eq:branches}
\begin{pmatrix}
D_\alpha \\ D_\beta
\end{pmatrix}^{(i+1)}
=\frac{1}{\xi_\beta - \xi_\alpha}
\begin{pmatrix}
\xi_\beta & -1 \\
-\xi_\alpha & 1
\end{pmatrix}
\begin{pmatrix}
D_0 \\ D_h
\end{pmatrix}^{(i)},
\end{equation}
where the superscripts indicate the layer in question.  It should be noted that $\vec{k}_{\alpha,\beta}$ and $\xi_{\alpha,\beta}$ are all functions of $\eta$, and it is through this parameter that the strain is incorporated.  Therefore, in \Equation{eq:branches}, care must be taken that the $\xi_{\alpha,\beta}$ are evaluated in the correct layer. We may do so by considering each layer as a separate crystal, with the beams $D_{o,h}^{(i)}$ exiting crystal $(i)$ and then entering crystal $(i+1)$ without change. It is then easy to see that the correct layer in which to evaluate these quantities is layer $(i+1)$.

\subsection{Incorporating Strain}

Strain is incorporated into the model through the deviation parameter $\eta$ (or more specifically, its real part $\eta_{r}$), which can be expressed as a function of the angular deviation from the Bragg angle $\Delta\theta$ (\Equation{eq:eta-definition}).  Strain affects this angular deviation in two ways: (i) rotating the planes (changing $\theta$) and (ii) altering the separation (changing $\bragg$).  The method presented here is similar to that of earlier work in Bragg geometry~\cite{speriosu81,speriosu84,wie86}, but is now presented for the Laue geometry case.

\begin{figure}
\begin{center}
\includegraphics[width=\columnwidth]{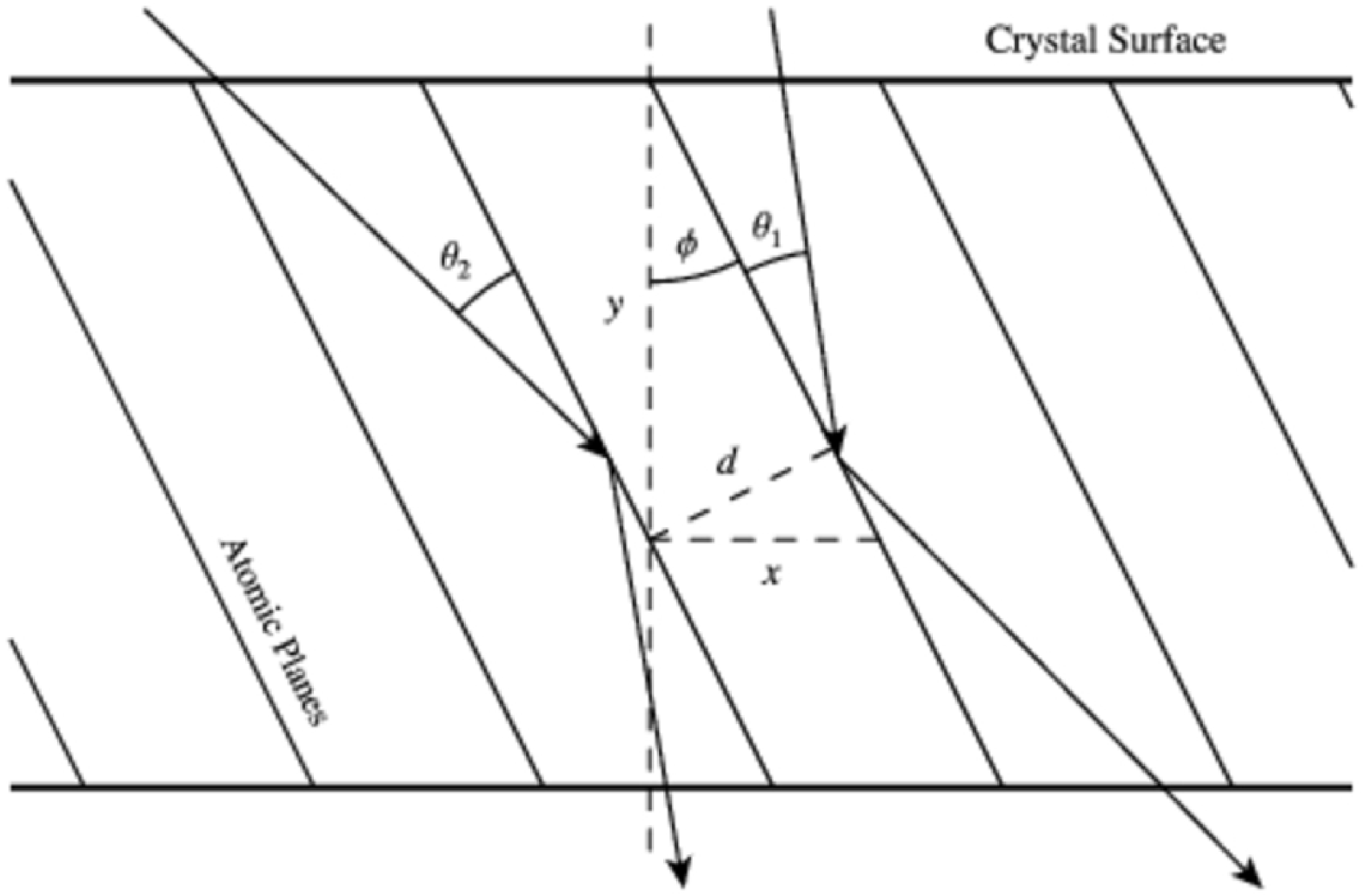}
\caption{Strained crystals. Strain ($\strain_\perp=\frac{\delta y}{y}$) alters the rotation ($\delta\phi$) and the separation ($\delta d$) of lattice planes in the crystal.  Note that there are two possible orientations for the x-ray beams.}
\label{fig:plane-rotation}
\end{center}
\end{figure}

From \Figure{fig:plane-rotation} we can formulate the following relations by taking differentials of the relationships between $d$, $x$, $y$ and $\phi$.
\begin{align}
\delta\phi&=-\strain_{\perp}\sin\phi\cos\phi,  \label{eq:delta-phi}\\
\frac{\delta d}{d}&=\strain_{\perp}\sin^2\phi, \label{eq:delta-d}
\end{align}
where $\strain_\perp=\frac{\delta y}{y}$ is the strain perpendicular to the surface.


The change in diffraction angle $\delta\theta$ is equal to  $\pm\delta\phi$ depending on the orientation of the incoming and outgoing beams ($\theta_1$ or $\theta_2$ in \Figure{fig:plane-rotation} for positive and negative $\phi$ respectively).  The change in diffraction angle due to the planar rotation is therefore
\begin{equation}
\delta\theta=\mp\strain_\perp\sin\phi\cos\phi.
\end{equation}


The change in the Bragg angle can be calculated using the differential form of Bragg's law, $$\delta\bragg=\frac{\delta d}{d}\tan\bragg,$$
combined with \Equation{eq:delta-d}, giving
\begin{equation}
-\delta\bragg=\strain_\perp\sin^2\phi\tan\bragg.
\end{equation}

We can now include both of these effects in the angular deviation from the Bragg angle to give the `effective misorientation'
\begin{equation}
(\theta-\bragg) \rightarrow (\theta-\bragg) + 
(\sin^2\phi\tan\bragg \mp \sin\phi\cos\phi)\strain_\perp.
\end{equation}

This can be used in every lamina, with the appropriate strain, in order to calculate the time-resolved rocking curves from the strained crystal.  In order for this lamellar approximation can be considered accurate, each lamina must be sufficiently thin that the change in the effective misorientation is small within it.  In practice this can be achieved by using a variable layer thickness, with a maximum allowed strain change across a layer.  The maximum strain change within a layer can then be reduced until the results obtained remain constant.  This was achieved with a maximum strain change of $1\times10^{-8}$ within any layer, for the geometry under consideration in this article.

\section{Strain Models}
\label{sec:strain-models}

As an example of this technique, we use the case of a Germanium single crystal irradiated by a femtosecond pulse of near infra-red laser radiation.  Experiments using Laue geometry time-resolved x-ray diffraction on such a sample have recently been carried out~\cite{decamp03}. This case has also been studied experimentally using ultrafast reflectivity \cite{auston74,auston75,chigarev00,tanaka95} and x-ray Bragg scattering techniques \cite{rose-petruck99,lindenberg00, reis01,siders99,cavalleri00,sokolowski-tinten01,cavalleri01}. 

In this experimental set-up, the short laser pulse quickly heats the surface at constant volume, generating a thermal stress.  This causes the surface to expand and, by Newton's third law, launches an acoustic pulse into the crystal.   In both cases considered here, the lateral size of the laser spot is assumed to be much greater than the laser absorption depth. As a result, the strain generated can be assumed to be one dimensional,\footnote{This is for times less than lateral extent of the spot divided by the speed of sound.  This corresponds to hundreds of nanoseconds for the experimental conditions modelled.} i.e.\ the atomic displacement only varying as a function of depth into the crystal.  

We will be comparing two models of the strain produced in this system.  The first model takes into account the time-scales of the processes by which the energy is transferred to the lattice.  The second is a simplified case with an analytic solution, introduced by Thomsen \textit{et al}~\cite{thomsen86}.  It assumes instantaneous transfer of energy from the laser into the lattice and no diffusion.
\begin{figure}
\begin{center}
\includegraphics[width=\columnwidth]{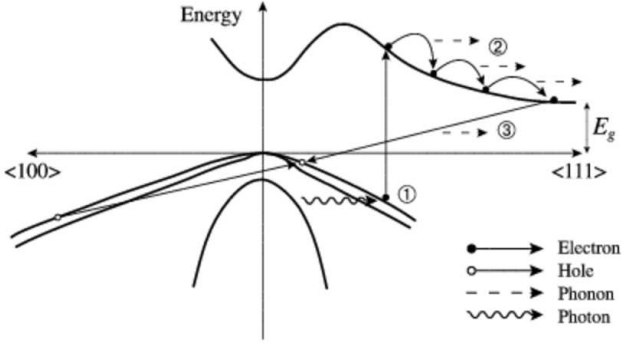}
\caption{Band structure of Germanium. The transfer of energy to the lattice involves: (1) Laser excitation. (2) Fast relaxation to band edges. (3) Auger recombination.}
\label{fig:simulation-processes}
\end{center}
\end{figure}

When a short laser pulse is incident on the crystal, the following energy transfer processes occur (see \Figure{fig:simulation-processes}):

\begin{enumerate}
\item The laser excites electrons from the valence band to the conduction band of the semiconductor, creating an electron-hole plasma.
\item The electrons and holes quickly relax to the band edges, transferring their energy to the lattice.  This energy transfer is assumed to be instantaneous on the time-scale considered ($\lesssim1\ps$).
\item The electrons and holes then recombine (by Auger recombination---an \textit{e-e-h} and \textit{e-h-h} process), transferring the rest of the energy to the lattice\cite{huldt71}. This takes significantly longer (typically $1\ns$ to $1\us$), allowing the carriers to diffuse appreciably into the crystal before recombining.
\end{enumerate}

The electron-hole plasma and the thermal phonons in the lattice both diffuse into the crystal bulk, with separate diffusion constants.  They both cause a change in the lattice spacing, through the deformation potential and thermal expansion respectively.

The laser energy is deposited in certain area of the crystal with a $1/\me$ absorption depth $\zeta$ and an absorbed fluence $Q$.  This gives initial conditions of the electron-hole plasma and the lattice temperature as
\begin{align}
\label{eq:initial-conditions}
n(z,t=0)&=\frac{Q}{E_{p}\zeta}\me^{-z/\zeta},\\
T(z,t=0)&=\frac{E_{p}-E_{g}}{C_{l}}n(z,t=0),
\end{align}
where $n$ is the electron-hole plasma density, $E_{p}$ is the energy of the laser photons, $E_{g}$ is the indirect band gap and $C_{l}$ is the lattice heat capacity per unit volume.

 The electron-hole plasma obeys a diffusion equation with a sink term for Auger recombination
\begin{equation}
\label{eq:plasma-diffusion}
\frac{\partial n}{\partial t}= D_{p} \frac{\partial^{2}n}{\partial z^{2}}- An^{3},
\end{equation}
where $D_{p}$ is the plasma diffusion constant and $A$ is the Auger recombination rate.  The energy from the Auger recombination is then transferred to the lattice, which obeys a diffusion equation with a corresponding source term
\begin{equation}
\label{eq:thermal-diffusion}
\frac{\partial T}{\partial t}= D_{t} \frac{\partial^{2}T}{\partial z^{2}}+ An^{3}\frac{E_{g}}{C_{l}},
\end{equation}
where $D_{t}$ is the thermal diffusion constant.

The equilibrium strain is then
\begin{equation}
\label{eq:equ-strain}
\strain_{e}(z,t)=\alpha_{t}T(z,t) + \alpha_{p}n(z,t),
\end{equation}
where $\alpha_{t}$ is the thermal expansivity ($=\Phi\beta$, $\Phi$ is a factor to take into account the 1D nature of the strain, $\beta$ is the linear expansion coefficient) and $\alpha_{p}$ is an electronic contribution to the strain associated with the deformation potential ($=\partial (\log a)/\partial n$, $a$ is the equilibrium lattice constant).  Any change in the equilibrium strain produces forward and backward propagating waves, which can be calculated by integrating \Equation{eq:equ-strain} as follows:
\begin{subequations}
\label{eq:fw-bw-waves}
\begin{equation}
\label{eq:forward-wave}
\strain_{+}(z,t)=-\frac{1}{2}\int_{0}^{t}\frac{\partial}{\partial t}\big[
\strain_{e}(z-vt',t-t')\big] \dif t',
\end{equation}\begin{equation}
\label{eq:backward-wave}
\strain_{-}(z,t)=-\frac{1}{2}\int_{0}^{t}\frac{\partial}{\partial t}\big[
\strain_{e}(z+vt',t-t')\big] \dif t',
\end{equation}
\end{subequations}
where $v$ is speed of longitudinal sound in the crystal.  The free-surface boundary condition of the stress, $\sigma_{33}(z=0,t)\equiv0$, is ensured by defining $$\strain_{e}(-z,t)\equiv -\strain_{e}(z,t),\quad z>0.$$  The total strain is then the sum of the equilibrium and the forward and backward going strain waves.
\begin{equation}
\label{eq:total-strain}
\strain(z,t)=\strain_{e}(z,t)+\strain_{+}(z,t)+\strain_{-}(z,t).
\end{equation}

\begin{figure}
\begin{center}
\includegraphics[width=\columnwidth]{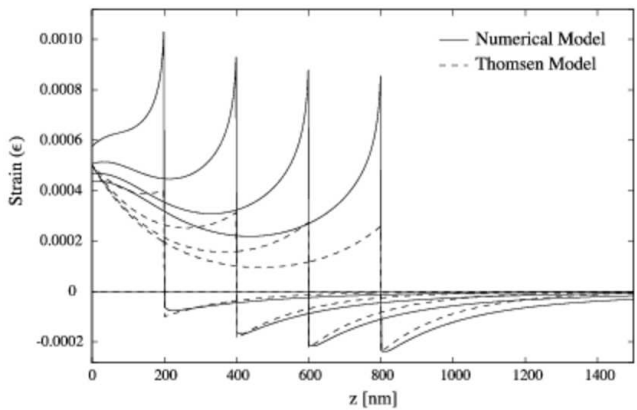}
\caption{Strain profile evolution---comparison of numerical and Thomsen models. This figure shows a time sequence of strain snapshots taken at $\sim40\ps$ intervals ($v=4915\mps$) for an absorbed fluence of $Q=4\fluence$. The physical parameters used are the literature values given in \Section{sec:results}.  The strain in the numerical model comprises a diffusing surface component and an asymmetric travelling wave.  The asymmetry is caused by the $e$-$h$ diffusion, the higher peak strain by the deformation potential, and the surface decay by a combination of thermal and $e$-$h$ diffusion.  The strain in the Thomsen model is made up of an exponential surface component and an anti-symmetric travelling wave. The travelling portions of the strain move into the crystal at the speed of sound. }
\label{fig:strain-comparison}
\end{center}
\end{figure}

The Thomsen strain model \cite{thomsen86} in its most simple form assumes instantaneous transfer of energy into the lattice and no diffusion.  In the time-scale of interest only a fraction $(E_{p}-E_{g})/E_{p}$ of the absorbed laser energy is transferred to the lattice (the rest remaining in the electron-hole plasma).  This gives a total strain of
\begin{equation}
\begin{split}
\label{eq:thomsen-strain}
\strain_{\text{Th}}(z,t)=&\frac{Q\Phi\beta(E_{p}-E_{g})}{\zeta C_{l} E_{p}}\left\{
\me^{-z/\zeta}\right.\\&\left.-
\frac{1}{2}\left[
\me^{-(z+vt)/\zeta}+\me^{-\abs{z-vt}/\zeta}\sgn{(z-vt)}\right]
\right\}.
\end{split}
\end{equation}
Note that this is the sum of a non-evolving exponential surface strain, a forward going wave $\mathcal{F}(z-vt)$ and a backward going wave $\mathcal{G}(z+vt)$.

The numerical model produces an identical strain to the above analytic formula, if there is no deformation potential or Auger recombination ($\alpha_{p}=0$ and $A=0$). A comparison of the strains produced by the numerical and Thomsen strain models is shown in \Figure{fig:strain-comparison}.

\section{Simulation}

\begin{figure}
\begin{center}
\includegraphics[width=\columnwidth]{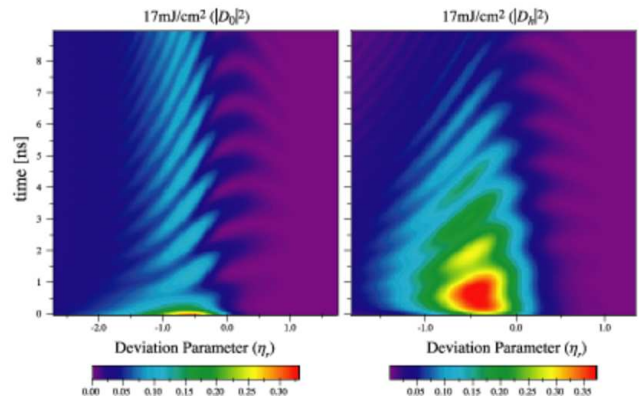}
\caption{Simulated time-resolved rocking curves. The $0$ and $h$ beam rocking curves plotted as a function of time for the numerical strain model with an absorbed pump fluence of $Q=17\fluence$.}
\label{fig:simulation-plots}
\end{center}
\end{figure}

The diffraction simulation works as follows:  Over a number of time-steps, the strain profile (analytic or numerical) is calculated.  For the numerical method, the diffusion equations (\ref{eq:plasma-diffusion} and \ref{eq:thermal-diffusion}) are solved using a Crank-Nicholson scheme, and the integrals in \Equation{eq:fw-bw-waves} are calculated by finite differencing then summing.  The x-rays are then propagated through the crystal over the $\eta_{r}$ range of interest.  This is done using the propagation matrices (\Section{sec:dynamical-theory}).  After the final layer, the amplitudes for the $0$ and $h$ beams are multiplied by their complex conjugates, giving the rocking curves (\Figure{fig:simulation-plots}).
The resulting time-resolved rocking curves are then integrated over $\eta_{r}$ and normalised such that the value for an unstrained crystal is unity, for comparison with experimental results.  

It is also possible to output the beam intensities through the bulk of the crystal: either as the $0$ and $h$ beams, showing Pendell\"osung oscillations~\cite{batterman64}; or as the $\alpha$ and $\beta$ branches, showing the transfer of energy between the two as the strain moves through the crystal (see \Figure{fig:alpha-beta}).

\section{Experiment}\label{sec:experiment}

\begin{figure}[bth]
\begin{center}
\includegraphics[width=\columnwidth]{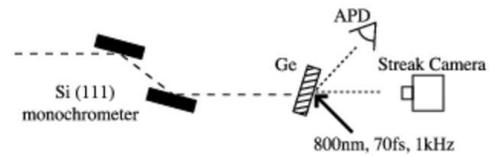}
\caption{Experimenal setup. The $h$ beam is collected by an avalanche photo-diode (APD) giving a time resolution of the synchrotron bunch length, the $0$ beam is resolved by a streak camera giving picosecond resolution.}
\label{fig:michigan-experiment-layout}
\end{center}
\end{figure}

The experimental data was taken at the 7-ID undulator beamline at the Advanced Photon Source.  The x-ray beam energy was $10\keV$ with a $1.4\times10^{-4}$ fractional energy spread (larger than the rocking curve width of the crystal) and negligible beam divergence.  The sample was a $280\um$ thick $(001)$ Germanium single crystal, oriented to diffract from the $20\bar{2}$ planes (\Figure{fig:michigan-experiment-layout}).  The x-ray beam was masked by tantalum slits giving a beam size on the crystal of $400\um \times 400\um$. The strain pulse was produced by exciting the output face of the crystal with sub-$100\fs$, $800\nm$ laser pulses. The diffraction simulation was set up to match these experimental conditions.

\section{Results and discussion}\label{sec:results}

\begin{figure*}[htbp]
\begin{center}
\includegraphics[width=\linewidth]{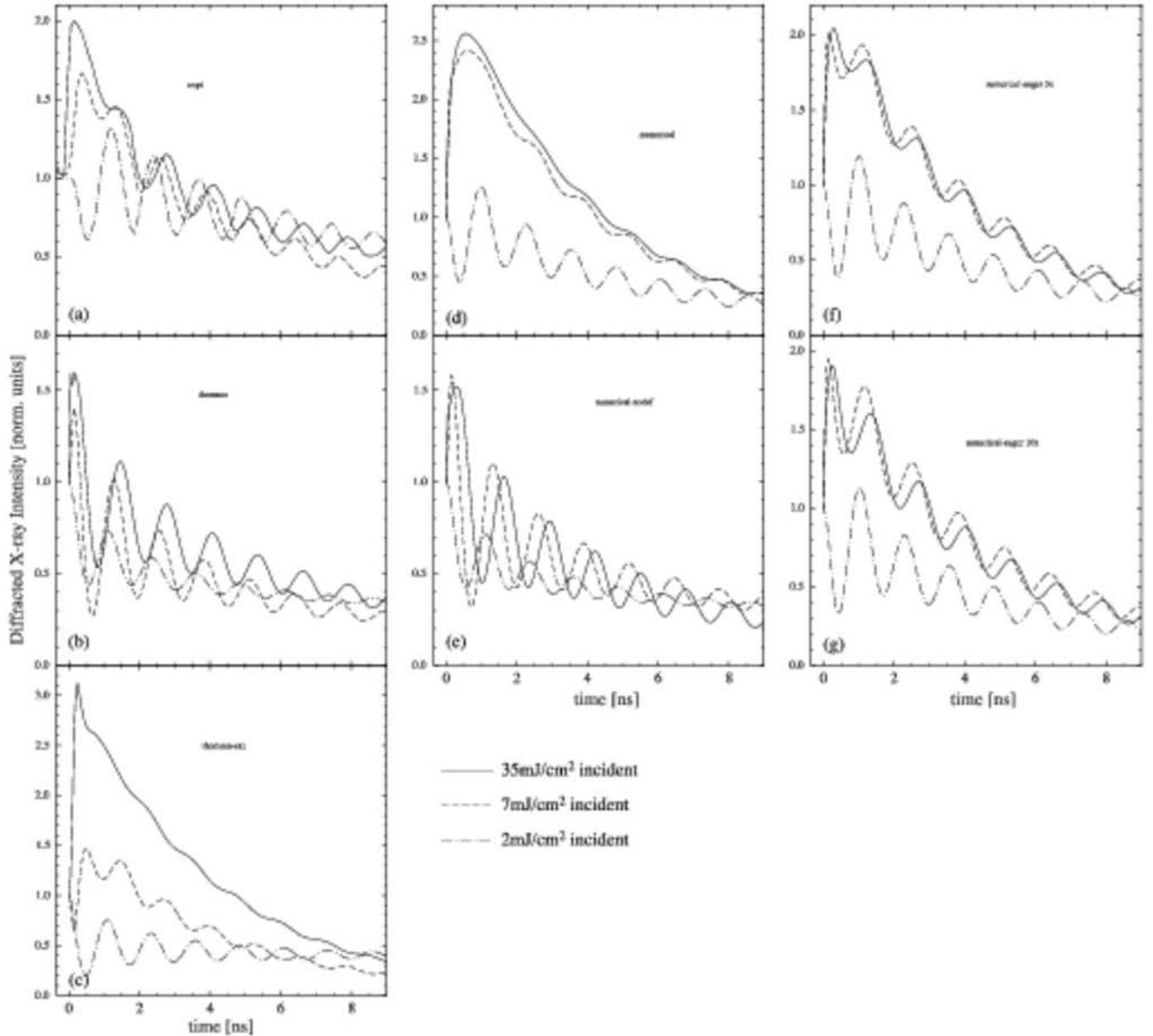}
\caption{Time-resolved integrated reflectivity. Plots are for diffracted beam ($\abs{D_{h}}^{2}$). (a) Experimental results. (b) Thomsen model with $\zeta=200\nm$. (c) Thomsen model with $\zeta=1\um$.  (d) Numerical model, $D_{t}=0.35\cm^{2}/\text{s}$, $D_{p}=65\cm^{2}/\text{s}$, $A=1.1\times 10^{-31}\cm^{6}/\text{s}$, $\alpha_{t}=10^{-5}\text{K}^{-1}$, $\alpha_{p}=1.3\times10^{-24}\cm^{3}$, $\zeta=200\nm$, $E_{g}=0.67\eV$, $C_{l}=1.7\,\text{J}/\text{K}\text{cm}^{3}$ and $v=4915\mps$. (e) Numerical model with no deformation potential, as (d) but with $\alpha_{p}=0$. (f) and (g) Numerical model with increased Auger recombination, as (d) but with $A=5.5\times 10^{-31}\cm^{6}/\text{s}$ and $11\times 10^{-31}\cm^{6}/\text{s}$ respectively.   Note that modelling assumes that 75\% of the incident laser energy is absorbed.}
\label{fig:results}
\end{center}
\end{figure*}

\begin{figure}
\begin{center}
\includegraphics[width=\columnwidth]{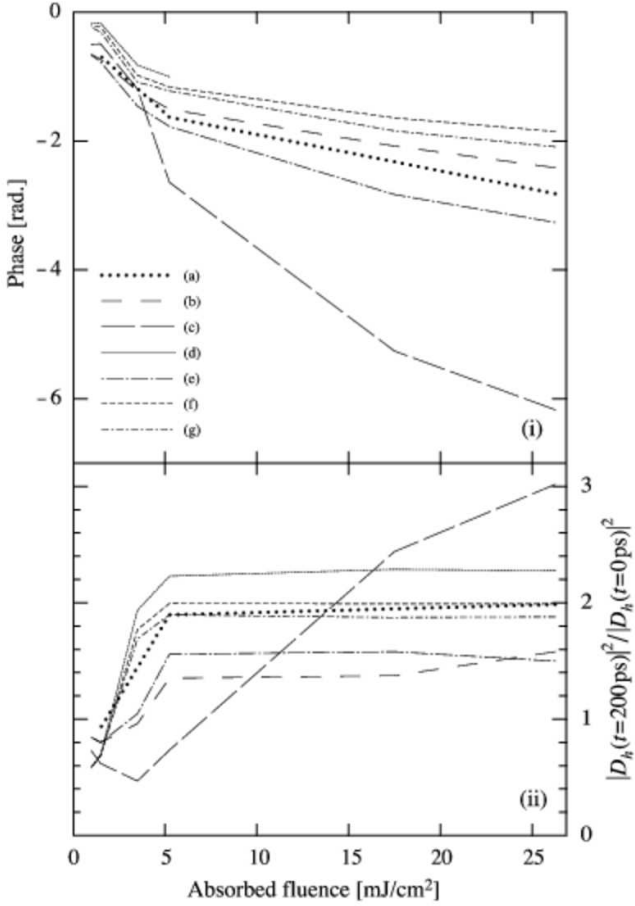}
\caption[Effects of changing the deformation potential and the plasma diffusion constant on (i) the phase of the oscillations and (ii) the initial rise or fall of $\abs{D_{h}}^{2}$.]{Effects of changing the deformation potential and the plasma diffusion constant on (i) the phase of the oscillations and (ii) the initial rise or fall of $\abs{D_{h}}^{2}$.  The phase is fitted to curves of the form shown in \Equation{eq:phase-fit}, with the data starting at $t=1\ns$ to ignore the initial rise. Fitted to: (a) Experimental data (assuming 75\% absorption). (b) Thomsen strain model, $\zeta=200\nm$. (c) Thomsen model, $\zeta=1\um$. (d) Numerical strain model, $\alpha_{t}=1\times10^{-5}\,\text{K}^{-1}$, $\alpha_{p}=1.3\times10^{-24}\cm^{3}$, $A=1.1\times 10^{-31}\cm^{6}/\text{s}$. (e) Numerical model with no deformation potential, as (d) but with $\alpha_{p}=0$. (f) and (g) Numerical model with increased Auger recombination, as (d) but with $A=5.5\times 10^{-31}\cm^{6}/\text{s}$ and $11\times 10^{-31}\cm^{6}/\text{s}$ respectively.}
\label{fig:results-phase-comparison}
\end{center}
\end{figure}

Data taken experimentally~\cite{decamp03} is shown in \Figure{fig:results}(a).  Oscillations are seen in the $h$ beam amplitude, with the time evolution of these varying with pump fluence.  The phase of the oscillations is calculated by fitting curves of the form
\begin{equation}\label{eq:phase-fit}
I(t)=\sum_{i}a_{i}\me^{-\frac{t}{b_{i}}}\sin(\omega_{i}t-\phi_{i}),
\end{equation}
starting at $t=1\ns$ (to ignore the initial rise).  The size of the initial rise or fall is measured by taking the normalised intensity at $t=200\ps$.  The phase $\phi$ and the initial rise or fall ($\abs{D_{h}(t=200\ps)}^{2}/\abs{D_{h}(t=0\ps)}^{2}$) of the measured and simulated reflectivity is shown in \Figure{fig:results-phase-comparison}. The curves for measured reflectivity assume $75\%$ of the laser energy is absorbed by the crystal.

Simulations were run with the Thomsen strain model, using literature values for physical parameters ($\zeta=200\nm$,  $E_{g}=0.67\eV$, $\Phi\beta=1\times10^{-5}\,\text{K}^{-1}$, $C_{l}=1.7\,\text{J}/\text{K}\cm^{3}$).  The results are shown in Figs.\ \ref{fig:results} and \ref{fig:results-phase-comparison}(b).  These match well with the phase change of the oscillations (to approx 10\%), but show a very different overall form and don't match the initial rise.  If the laser absorption depth $\zeta$ is increased to $1\um$ (Figs.\ \ref{fig:results} and \ref{fig:results-phase-comparison}(c)), the overall form of the higher fluence curves shows a better match. However, the phase of the oscillations doesn't show the expected behaviour.

Using the more detailed numerical model, again with literature values for physical parameters ($\zeta=200\nm$, $\alpha_{t}=1\times10^{-5}\,\text{K}^{-1}$, $\alpha_{p}=1.3\times10^{-24}\cm^{3}$, $A=1.1\times 10^{-31}\cm^{6}/\text{s}$, $D_{t}=0.35\cm^{2}/\text{s}$, $D_{p}=65\cm^{2}/\text{s}$), the curves in Figs.\ \ref{fig:results} and \ref{fig:results-phase-comparison}(d) were calculated.  These match the fluence dependence of the experiment in the overall form of the curves and the initial rise.  However, the phase of the oscillations only decreases at about two-thirds the rate of the experimentally measured phase.  At the highest fluences the oscillations in $D_{h}$ are of such small magnitude that it was impossible to fit the curves numerically.

If the deformation potential is taken out of the numerical model (by setting $\alpha_{p}=0$), the correct fluence dependence of the phase is obtained (to approximately 10\%), but the behaviour at early times (ie. the initial rise or fall) is no longer well matched (Figs.\ \ref{fig:results} and \ref{fig:results-phase-comparison}(e)).

These last two observations would imply that there is a direct electronic contribution to the strain (through the deformation potential) at earlier times, which is greatly reduced at later times. One possible mechanism for this is that the Auger recombination rate is larger than expected.  Simulations were run with the recombination rate increased five-fold and ten-fold ($A=5.5\times 10^{-31}\cm^{6}/\text{s}\;\text{and}\;11\times 10^{-31}\cm^{6}/\text{s}$ respectively).  The resulting curves (Figs. \ref{fig:results} and \ref{fig:results-phase-comparison}, (f) and (g) respectively) show good agreement with the experimental phase, and a reasonable agreement with early time behaviour.

We were unable to find an excellent match with any combination of parameters, but this particular set gave the best of those tried.  The exact mechanism for a larger than expected Auger recombination rate is unknown.  One possible explanation is that the higher temperature of the lattice at early times activates other possible recombination pathways (leading to a temperature dependence to the recombination rate). It is thought that the main reason for the recombination rate being so low in Germanium is that it requires phonon activation~\cite{huldt74}.  So, at the higher temperatures existing at early times, this could cause faster than expected recombination.    More detailed modelling of the strain, taking this possible non-linearity into account, could lead to better reproduction of the experimental data.

\begin{figure}
\begin{center}
\includegraphics[width=\columnwidth]{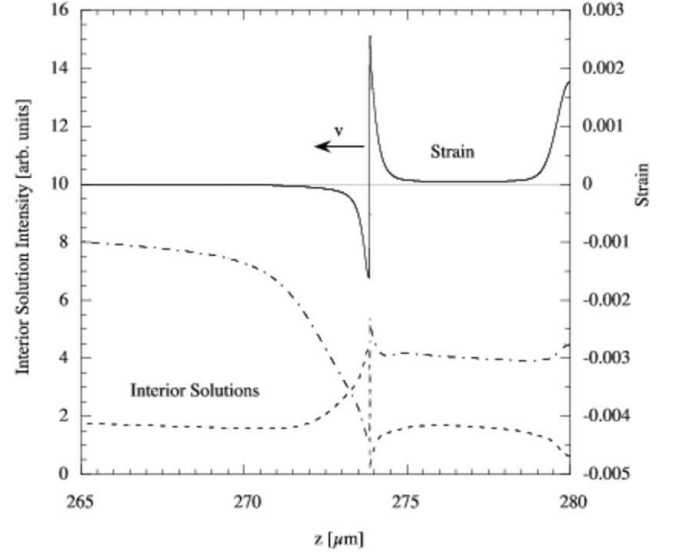}
\caption{The $\eta$-integrated interior solutions $\abs{D_{\alpha}}^{2}$ (dot-dashed line) and $\abs{D_{\beta}}^{2}$ (dashed line) as a function of depth into the crystal at $1.25\ns$ after laser excitation, for an absorbed pump fluence of $Q=17\fluence$. Also shown is the strain distribution (solid line).  Strain front is moving from right to left.  X-ray beams exit crystal on right.}
\label{fig:alpha-beta}
\end{center}
\end{figure}

In an earlier paper~\cite{decamp03}, the authors posited that the basic physics behind the oscillations visible in the $\eta$-integrated rocking curves (\Figure{fig:results}) was due energy being transferred from the $\alpha$ branch to the strongly absorbed $\beta$ branch at a disturbance in the lattice.  Using the diffraction simulation it is possible to look at how the strain transfers this energy between the solutions. \Figure{fig:alpha-beta} shows that energy is transferred from the $\alpha$ branch to the $\beta$ branch when the strain gradient is negative, and is then quickly transferred back at the positive strain discontinuity.  This energy transfer effect is reversed for the opposite asymmetry ($\bar{2}02$).

\begin{figure}
\begin{center}
\includegraphics[width=\columnwidth]{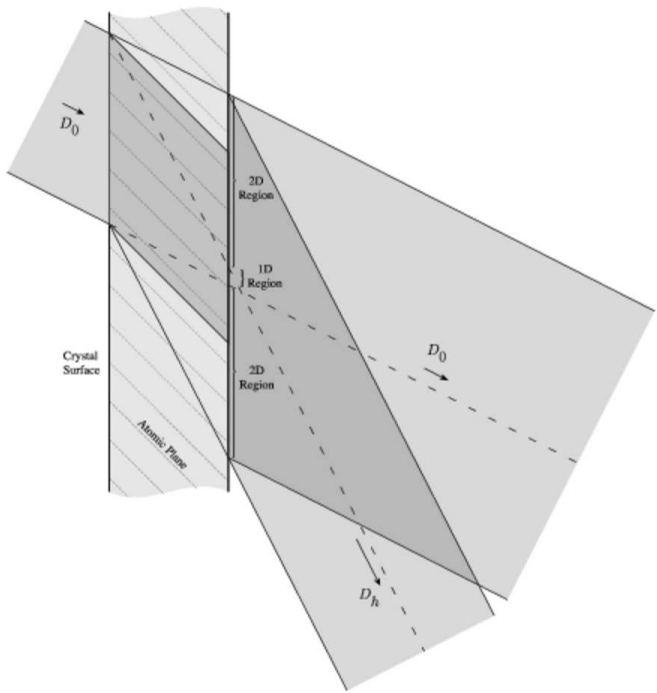}
\caption{2D Region (drawn to scale). The incoming $D_{0}$ beam has a width of $400\um$, the crystal is $280\um$ thick. In the `2D region' shown, the exit surface of the crystal is not influenced by the whole incoming beam.}
\label{fig:2d-region}
\end{center}
\end{figure}

\begin{figure}
\begin{center}
\includegraphics[width=\columnwidth]{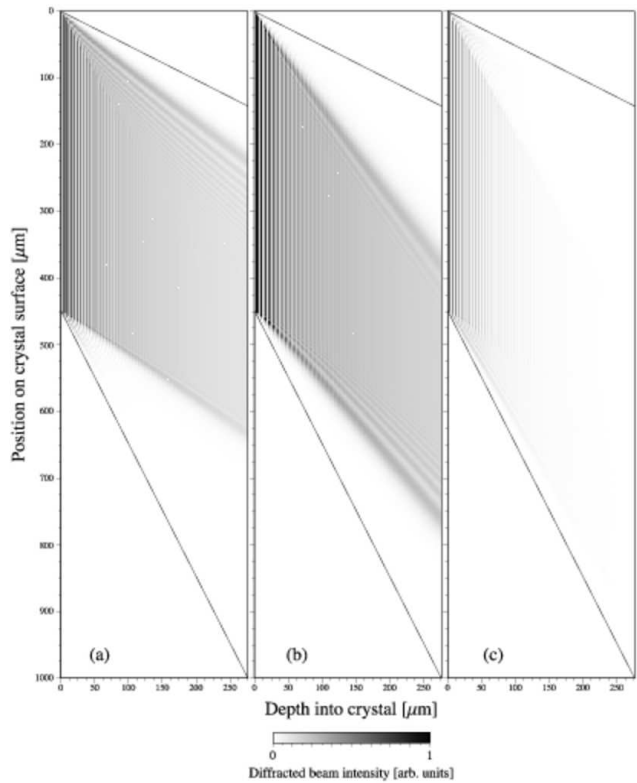}
\caption{2D simulation results (drawn to scale). $\abs{D_{h}}^{2}$ within the crystal for (a) $\eta_{r}=-1$, (b) $\eta_{r}=0$, (c) $\eta_{r}=+1$.}
\label{fig:2d-results}
\end{center}
\end{figure}

It might initially be expected that analysis of these experimental results would need a 2D diffraction model. This is because, as shown in \Figure{fig:2d-region}, for this experiment only a small fraction of the exit surface is influenced by the entire incoming x-ray beam---the area of influence being the inverted Borrmann triangle from the exit point\cite{authier01}. The 2D Takagi-Taupin equations normally reduce to a case only dependent on the depth into the crystal for a one dimensional strain, as in this case\cite{takagi62}. However, this also has the implicit assumption that the lateral extent of the x-ray beam is large enough for its size to be unimportant.   2D simulations were run, using the `half-step derivative' numerical solution \cite{authier68a}.  Sample results are shown in \Figure{fig:2d-results}.  For comparison with the 1D results, the intensity distribution was summed over the exit surface of the crystal, and then over the rocking curve.  The time-resolved integrated reflectivities obtained only differed by $0.25\%$ (RMS) from the 1D diffraction model.  The reason for this is that close to the rocking curve peak, the $D_{0,h}$ beams are not `eigensolutions' inside the crystal---the $D_{\alpha,\beta}$ solutions are, due to the strong coupling between the beams.  We would therefore only expect differences to become apparent in the rocking curve wings, where the coupling is less strong.  In the experimental conditions, where the data obtained is the $\eta$-integrated rocking curve, the great majority of the signal comes from the rocking curve peak, completely swamping any small changes in the wings.

\section{Conclusion}

We have presented a method for the calculation of time-resolved rocking curves for X-ray diffraction in Laue (transmission) geometry. Such diffraction studies are an important technique for the study of coherent strain in crystals beyond the extinction depth.  Even without the benefit of rocking curve resolution of the diffracted x-ray beams, it is possible to obtain valuable information about the form of the strain profile in the crystal, and to observe the effects of the mechanisms of the ultrafast energy transfer processes at work.  As there is no method to analytically compute the strain in a crystal from a time-resolved rocking curve, numerical simulations are key for the understanding of experimental results obtained.

\begin{acknowledgments}

The experimental work was conducted at the MHATT-CAT insertion device beam-line at the Advanced Photon Source and was supported in part by the US Department of Energy, Grants No. DE-FG02-99ER45743 and No. DE-FG02-00ER15031, by the AFOSR under contract F49620-00-1-0328 through the MURI program and from the NSF FOCUS physics frontier center.  Use of the Advanced Photon Source was supported by the US Department of Energy Basic Energy Sciences, Office of Energy Research under contract No. W-21-109-Eng-38. SF acknowledges the financial support of Science Foundation Ireland. BL acknowledges the financial support of EPSRC and AWE.

\end{acknowledgments}

\bibliography{paper}

\end{document}